# Massively Parallel Processor Architectures for Resource-aware Computing

Vahid Lari, Alexandru Tanase, Frank Hannig, and Jürgen Teich
Hardware/Software Co-Design, Department of Computer Science
Friedrich-Alexander University Erlangen-Nürnberg (FAU), Germany
{vahid.lari, alexandru-petru.tanase, hannig, teich}@cs.fau.de

*Abstract*—We present a class of massively parallel processor architectures called invasive tightly coupled processor arrays (TCPAs). The presented processor class is a highly parameterizable template, which can be tailored before runtime to fulfill costumers' requirements such as performance, area cost, and energy efficiency. These programmable accelerators are well suited for domain-specific computing from the areas of signal, image, and video processing as well as other streaming processing applications. To overcome future scaling issues (e. g., power consumption, reliability, resource management, as well as application parallelization and mapping), TCPAs are inherently designed in a way to support self-adaptivity and resource awareness at hardware level. Here, we follow a recently introduced resource-aware parallel computing paradigm called invasive computing where an application can dynamically claim, execute, and release resources. Furthermore, we show how invasive computing can be used as an enabler for power management. Finally, we will introduce ideas on how to realize fault-tolerant loop execution on such massively parallel architectures through employing on-demand spatial redundancies at the processor array level.

## I. INTRODUCTION

The steady miniaturization of feature sizes allows to create increasingly complex Multi-Processor System-on-Chip (MPSoC) architectures but raises also numerous questions. These challenges include imperfections and unreliability of the devices as well as scalability problems of the architectures, as for instance, how an optimal communication topology or memory architecture should look like. The situation is even more severe with respect to power consumption because chips can handle only a limited power budget—but technology shrinking leads also to higher energy densities continuously. As a consequence, the potentially available chip area might not be fully utilized or at least not simultaneously. These phenomena are also known as *power wall* and *utilization wall* [1]. Other scalability issues, caused by the sheer complexity of exponential growth, are related to resource management as well as parallelization and mapping approaches. This leads to the following conclusion: Future systems will only scale if the mapping and runtime methods will considerably improve—this reasoning holds for both embedded and portable devices such as smartphones and tablets as well as large scale systems as used for high-performance computing. Customization and heterogeneity in the form of domain-specific components such as accelerators are the key to success for future performance gains [2].

Furthermore, such a high integration density will lead to more and more vulnerability of the circuits to malfunction due to thermal effects, circuitry wear-outs, or even external cosmic radiations. Therefore, the protection of multi-core systems against faults has gained intensive research interests during last decades. But the situation turns to be even more severe, when considering massively parallel architectures, with hundreds to thousands of resources, all working with real-time requirements.

As a remedy, we present a domain-specific class of massively parallel processor architectures called invasive tightly coupled processor arrays (TCPA), which offer built-in and scalable resource management. The term "invasive" stems from a novel paradigm called *invasive computing* [3], for designing and programming future massively parallel computing systems (e. g., heterogeneous MPSoCs). The main idea and novelty of invasive computing is to introduce resource-aware programming support in the sense that a given application gets the ability to explore and dynamically spread its computations to other processors in a phase called *invasion*, and then to execute code segments with a high degree of parallelism, based on the region of claimed resources on the MPSoC. Afterward, once the application terminates or if the degree of parallelism should be decreased, it may enter a *retreat* phase, deallocates resources and resumes execution again, for example, sequentially on a single processor.

TCPAs consist of an array of tightly coupled light-weight processor elements [4]. Such architectures are well suited as domain-specific companions in an MPSoC for acceleration of loop programs from digital signal processing and multi-media applications. Typically, an application is statically mapped in a spatially or temporally partitioned manner on such array processors. To overcome this rigidity, our processor arrays support at hardware-level the ideas of invasive computing such as cycle-wise invasion of processor resources and multiple parallel hardware-supported 1D and 2D invasion strategies. For this purpose, we integrated special hardware controllers in each Processor Element (PE) of a TCPA to enable extremely fast and decentralized resource allocation [5]. Additionally, these *invasion controllers* (*i*Ctrls) we show to enable hierarchical power management in TCPAs, see [6].

In our ongoing research, we aim to enhance this capability to support fault tolerance through invading duplicated/triplicated redundant regions, based on application needs for reliability as well as the transient fault rate in the system. As a research goal, we study the effects of transient faults on processor arrays based on their occurrence location, and accordingly investigate different approaches for placing hardware/software voters to check the correctness of redundant loop executions. In addition, in order to achieve a trade-off between flexibility, timing and hardware cost, hardware voters will be implemented as special functional units within PEs, which can be programmed to vote over any inputs, outputs, or internal registers of PEs.

The rest of the paper proceeds as follows: In the Section II, we discuss related work. Section III addresses the architecture of TCPAs. Section IV describes the incorporation of invasive computing in TCPAs and resulting options for power management. Approaches for fault-tolerant loop execution on TCPAs are presented in Section V. After considering final remarks of our proposed approach, Section VI concludes.

## II. Related Work

MPSoC architectures have a high degree of parallelism, offer high performance and might be highly versatile in comparison to single core processors. Examples of recent multi- and many-core architectures include the Xeon Phi coprocessor series with more than 60 cores, Picochip's PC-200 series [7] with 200–300 cores per device or Tilera's TILEPro 32-bit processor family with up to 72 VLIW processor cores[1].

Managing and supervising a huge amount of resources in future architectures, if performed completely centralized, may become a major system's performance bottleneck, and thus, current approaches may not scale any longer. Therefore, there is a strong trend toward dynamic exploitation of the available level of parallelism in many-core architectures based on application requirements. For instance, in the TRIPS project [8], an array of small processors is used for the flexible allocation of resources dynamically to different types of concurrency, ranging from running a single thread on a logical processor composed of many distributed cores to running many threads on separate physical cores.

The existing methods are relatively rigid since they either have to know the number of available resources at compile time, or the considered architecture is controlled centrally and does not provide any hardware support for resource management and workload distribution. In order to tackle these problems, we present a distributed hardware architecture for the resource management in TCPAs in Section IV, and show that these concepts scale better than centralized resource management approaches. But, before it, we generally introduce the architectural properties of TCPAs in the next section.

## III. Architecture of Tightly Coupled Processor Arrays

A TCPA [9], [4] is a highly parameterizable architecture template, and thus offers a high degree of flexibility. Some of its parameters have to be defined at synthesis time, whereas other parameters can be reconfigured at runtime. This architecture may be used as an accelerator for compute-intensive loop kernels in MPSoCs. As shown in Fig. 1, the heart of the accelerator comprises a massively parallel array of tightly coupled processor elements (PEs); complemented by peripheral components such as I/O buffers. A PE itself is again a highly parameterizable component with a VLIW (Very Long Instruction Word) structure (see bottom part of Fig. 1). Here, different types and numbers of functional units (e.g., adders, multipliers, shifters, logical operations) can be instantiated as separate functional units, which can work in parallel. The size of the instruction memory and register file is as well parameterizable. The register file consists of four different types of registers for the data as well as the control

[1]Tilera Corporation, http://www.tilera.com

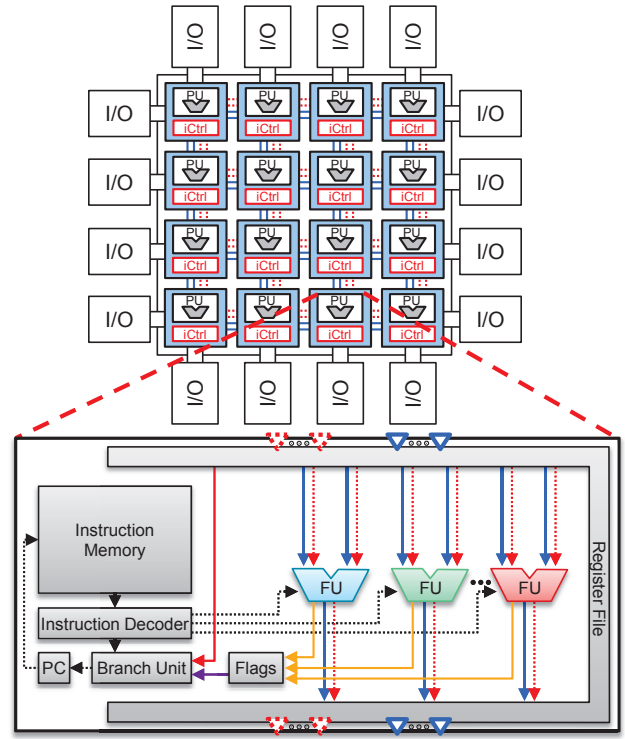

Fig. 1: A 4x4 tightly coupled processor array with I/O buffers surrounding the array, as well as a sketch of a single VLIW processing element.

path, i.e., input ports, output ports, general purpose registers, and rotating registers.

The array may consist of heterogeneous PEs. For instance, some of the PEs at the borders might include extra functionality for the purpose of address generation. However, in the rest of the paper, we consider a homogeneous array, which is augmented by dedicated address generation units. The PEs in the array are interconnected by a circuit-switched mesh-like interconnect with a very low latency, which allows data produced in one PE to be used already in the next cycle by a neighboring PE. An interconnect wrapper encapsulates each PE and is used to describe and parameterize the capabilities of switching in the network. The wrappers are arranged in a grid fashion and may be customized at compile-time to have multiple input/output ports in the four directions, i.e., north, east, south, and west. Using these wrappers, different topologies between the PEs like grid and other systolic topologies, but also (higher dimensional) topologies such as torus or 4-D hypercube can be implemented and changed dynamically.

Two different networks, one for data and one for control signals, can be defined by their data width and number of dedicated channels in each direction. For instance, two 16-bit channels and one 1-bit channel might be chosen as data and control network, respectively. Note that the data and control path width for the other architectural components such as functional units and registers is deduced from the selected channel bit widths. The processor array is surrounded by a structure of I/O buffers. These buffers can be configured to operate as FIFOs or normal random access memories (RAM). Furthermore, based on application requirements, multiple buffer banks could be concatenated together to form a larger memory for an application [10].

## IV. INVASIVE COMPUTING AND POWER MANAGEMENT

Invasive computing is a recently introduced resource-aware parallel computing paradigm that has been conceived to overcome the resource scalability problems in future many-core architectures [3]. This paradigm defines three phases of execution for each application, i.e., *invade* phase in which the application requests for its required resources as well as functional and non-functional execution properties. The acquired resources are loaded with binary of a parallel program, followed by execution of the program during an *infect* phase. When the application execution is completed or if the degree of parallelism should be reduced, the captured resources (or a portion of them) will be released by issuing a *retreat* request.

As outlined before, TCPAs are envisioned as accelerators in a heterogeneous tiled architecture suitable for running computationally intensive kernels. In this sense, a TCPA could be either an accelerator connected to RISC cores through data caches, or a shared bus, or even being a separate tile in a tiled architecture [4]. When an application, running on a RISC core, reaches an execution point requiring a high degree of parallelism, it requests for acceleration, i. e., a TCPA. This request reaches the TCPA architecture through the operating system. A dedicated control processor evaluates the availability of resources and places an invasion request on the processor array. In order to place an invasion in a processor array, first a suitable initiation point, or a so-called invasion seed, has to be selected. An intelligent selection of the seed invasion not only increases the chance of acquiring the required resources but also different system-level optimization objectives such as load and temperature might be balanced. The invasion seed is chosen among one of the PEs at the border of the array, where a direct access to the I/O buffer structure is guaranteed. The candidate invasion seeds are interfaced to the control processor. The number of interfaced seed PEs represents the capacity of a TCPA for running concurrent applications. This capacity is limited by the hardware cost, whereas for each concurrent application a full set of peripheral components is needed.

When the control processor receives an *invade* request, it chooses the best initiation point by checking the status of the *i*Ctrls. On the processor array this process takes place in a distributed fashion, where it starts from an initial element (e. g., corners of the array) by asking its neighboring resources about their availability by sending invasion signals over a dedicated control network. This process is continued by the neighboring resources in a cycle-by-cycle fashion until the required number of resources is reserved or no further resources can be claimed. As soon as the application's requirements (the number of needed PEs) are fulfilled, the invaded PEs start sending back confirmation signals, indicating information about the invaded PEs, such as the number of invaded PEs and their location. This process is performed in the reverse direction of the invasion, starting from the last PE in the invaded domain to the PE that has initiated the invasion. After subsequently loading the configuration into the captured PEs and completing the execution of the application, the initiator PE will issue a retreat signal to its captured neighbors. This signal is locally propagated through the captured PEs following the same path that the invade signal has paved until all captured PEs by the application are signaled to be released, which is again confirmed by a wave of signals from the last PE to the initiator PE.

Based on many application requirements for 1-D and 2-D image, signal processing, and other compute-intensive loop specifications, two major invasion strategies are proposed: *linear invasion* and *rectangular invasion* in [5]. Linear invasion, which targets at capturing a linearly connected chain of PEs, is suitable for types of one-dimensional digital signal processing applications such as FIR filters and rectangular invasions are suitable for two-dimensional applications such as image processing kernels, e.g., optical flow algorithm [11].

In order to support the propagation of invasion signals, each processor element of a many-core architecture must be equipped with an invasion controller (*i*Ctrl) [5]. This controller should be able to send invasion-related signals in a fast cycle-based manner and at the minimal hardware cost. In order to make a trade-off between the flexibility and performance, we considered the following designs for the invasion controller:

- Hard-wired FSM-based controllers that implement just a single invasion strategy,
- Programmable controllers that are flexible and can implement different invasion strategies.

A programmable invasion controller can be easily reprogrammed for a wide range of invasion strategies, so that it can be adapted for different application requirements. A solution based on a dedicated Finite State Machine (FSM) allows typically faster resource exploration, but is rigid and inflexible. Although the programmable controllers offer more flexibility for implementing different strategies compared to the FSM-based controllers, they suffer from higher invasion latencies. With a dedicated invasion controller, we gain speedups of 2.6 up to 45 compared against a purely software-based implementation for resource management on the control processor [5].

The resulting architecture of TCPAs minimizes the overhead of control flow, memory access, as well as data transmissions by inheriting the dynamicity and self-adaptiveness of invasive TCPAs. The energy consumption can be optimized by dynamically powering-off the idle regions in the array at retreat time (details follow in the subsequent section). This feature especially helps when using this architecture as a hardware accelerator inside portable devices, where battery life is critical.

### A. Invasive Computing as an Enabler for Power Management

Resource-aware computing shows its importance and advantages when targeting many-core architectures consisting of tens to thousands of processing elements. Such a great number of computational resources allows to support very high levels of parallelism but on the other side may also cause a high power consumption. One traditional way to decrease the overall power dissipation on a chip is to decrease the amount of static power by powering off unused resources [12]. In the context of invasive computing, we therefore exploited invasion requests to wake up processors and retreat requests to shut down the processors in order to save power. As these invasion and retreat requests are initiated by each application, the architecture itself adopts to the application requirements in terms of power needs. During the invasion phase, two different kinds of power domains are considered: *processing element power domains* and *invasion controller power domains*. These domains are controlled hierarchically, based on the system utilization which is in turn controlled by the invasion controllers (see Fig. 2). Whenever a PE receives an invade signal, its *i*Ctrl is first powered on, subsequently when the invasion is confirmed by a claim signal, the processing unit is turned on (making the PE ready to start application execution). Similarly, by receiving a retreat signal, both components are turned off again.

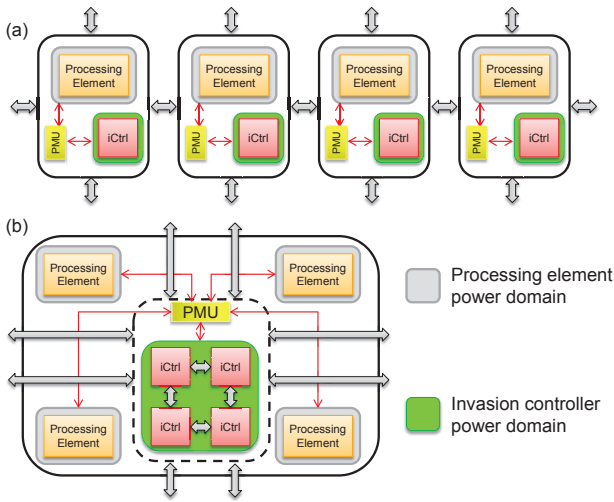

Fig. 2: Different designs for grouping invasion controllers into one power domain [6]. (a) Invasion controller power domains controlling the power state of a single invasion controller; (b) An invasion controller power domain controlling the power state of four invasion controllers belonging to four processor elements.

Power gating of individual invasion controllers may reduce the power consumption of the MPSoC but at the cost of timing overhead of power switching delays. In [6], we therefore studied the effects of grouping multiple invasion controllers in the same power domain. Such grouping mechanisms may reduce the hardware cost for power gating, yet sacrificing the granularity of power gating capabilities. The finer the granularity for the power control, the more power we may save. In contrast, grouping more invasion controllers together will reduce the timing overhead that is needed for power switching during both invasion and retreat phases. Fig. 2 shows different proposed example architectures for grouping the invasion controllers. Experimental results show that up to 70 % of the total energy consumption of a processor array may be saved for selected applications and different resource utilization. In addition, we presented a model for energy consumption based on the size of the invasion controller power domains in [6]. Notably, the estimation error of the presented models is less than 3.6 % in average when compared to simulation results.

## V. Fault Tolerance on Demand

One consequence of the device miniaturization and reduction of operating voltages is the increase of fault and failure rates that menace the correct functionality of computer systems. There is a rich literature on approaches for protecting systems against faults. Fault tolerance in a digital system may be achieved through redundancy in *hardware*, *software*, *information*, and/or *computations*. There are many works trying to protect systems against faults through hardware approaches such as redundant combinational circuits, self-checking circuits, or logic-level circuit hardening [13], [14]. However, pure hardware solutions typically ignore knowledge about a running application. Therefore, many researchers have investigated software-based fault tolerance through compiler techniques [15], [16].

Concerning Coarse-Grained Reconfigurable Architectures (CGRAs), the architecture itself manifests a natural redundancy at PE level in addition to the instruction level in case of superscalar or VLIW structured PEs. However, compared to standard general purpose processors, there are few works dealing with fault tolerance on such architectures: Schweizer et al. [17] propose a hardware-based approach for CGRAs with minimum overhead by using spare functional units for replication. Here, an error handling hardware component called Flexible Error Handling Module (FEHM) is integrated into a PE. FEHM supports Dual Modular Redundancy (DMR) and Triple Modular Redundancy (TMR) at either functional unit (FU) level or even PE level within a set of PEs clustered together. However, with the intensity of data computations on such architectures, the number of functional units within a PE might not be sufficient to explore such redundancy.

Software-based fault tolerance approaches for CGRAs are considered in the following approaches: [18] proposes instruction duplication in order to detect soft errors in VLIW datapaths. Here, the compiler determines the instruction schedule by filing empty execution slots with duplicate instructions. In the same way, Lee et al. [19] propose a software-based dual/triple replication of programs on a CGRA. In order to reduce the performance degradation, the recent work [20] shows how to reduce the number of validation points by applying software-based voters only to memory store operations. In summary, most of the mentioned work considers either pure software approaches for fault tolerance that may lead to considerable timing overheads in execution. Alternatively, hardware redundancy approaches might be too costly in terms of additional hardware complexity.

Therefore, providing a fault tolerance solution for processor arrays on demand and (a) with little timing and area overheads, (b) aware of the running application and sensitivity of system towards errors, and (c) on demand by exploiting hardware/software trade-offs offered by TCPAs is our major focus. Of particular concern here are compute-intensive loop programs and the protection of their parallel execution against soft errors by employing different types of redundancies adaptively, according to the system status. This includes the exploitation of the reconfigurability of TCPAs to implement dual- and triple-replicated implementations through invasive computing and on-the-fly. Finally, concepts for hardware and software voting need to be investigated that shall have no or only a minor impact on the performance of parallel loop execution.

Therefore, we study different aspects of software/hardware-based fault tolerance on TCPAs (see Fig. 3). Here, an application programmer may request for reliability (e.g. tolerance of single bit errors) while constructing an invade request, then based on the system vulnerability to faults, different levels of redundancies might be realized at array level. In case of memories and input buffers of TCPAs, communication media and the other components of the MPSoC, we assume the use of existing well-known fault tolerance approaches in the literature. As the first study, we will investigate at compiler-level the effect of replication and then voter placement on loop variables or final outputs (mentioned as software approaches in Fig. 3). A compiler-assisted approach to safe (reliable) loop processing for invasive processor array architectures shall be developed and its timing overhead shall be evaluated. We propose to exploit the paradigm of invasive computing to naturally claim not a single, but a dual- or triple-replicated processor array to implement DMR and TMR on demand of an invading loop application. Based on a user requirement whether errors shall be either only detected or also corrected and whether the corresponding DMR and TMR scheme required shall be able to handle single or even multiple errors occurring in the computation of loops, the compiler shall assist in claiming the required spare arrays and in providing the proper loop transformations of a given loop program to also compute the required voting operations. In order to exploit as much parallelism as possible, compiler transformations like loop replication (see Fig. 4(a)) as well as introduction and scheduling of voting variables and corresponding operations, will be automatically handled by the compiler. Moreover, we will tackle the problem of selectivity, i.e., which loop variables (e.g., all left-hand side variables, only variables output to neighbor processors, etc.) have to be checked and how often (e.g., each loop iteration). For example in Fig. 4(a), the compiler has transformed the initial iteration space of the one-dimensional nested loop for TMR. Here, the loop body is replicated three times for mapping the code onto the three claimed processor arrays (two extra arrays claimed automatically by the compiler) and added the extra voting variables and corresponding operations. Of course, placing more voting operations comes at higher costs in terms of hardware and timing overhead, but leads to better capabilities in tolerating multiple faults and fault isolation. As Fig. 4 suggests, the voting could be implemented both at software or hardware level. The software-based voting comes with significant timing overhead, specially for architectures like TCPAs that are more suitable for compute-intensive kernels rather that control-intensive ones [20]. Fig. 4(b) and (d) show two examples in which the voting in both cases are performed on the intermediate variables as well as loop outputs on either middle PEs or all replicas, respectively. In case of hardware voting, the PEs should be equipped with proper voter component. Here, our contribution is to have a *voter functional unit* that could be programmed at software level and vote/compare any member of register file (PEs marked with red color in Fig. 4(a) and (c)). In this sense, the flexibility of software voting is employed and at the same time, the timing overhead is also reduced compared with the software approaches. Of course such a timing improvement comes with hardware cost which should be also evaluated and optimized by placing such capabilities in a selective set of PEs (see Fig. 3). The overall approach should be transparent to application programmers in a way all data propagation and voting operations should be inserted into generated codes automatically.

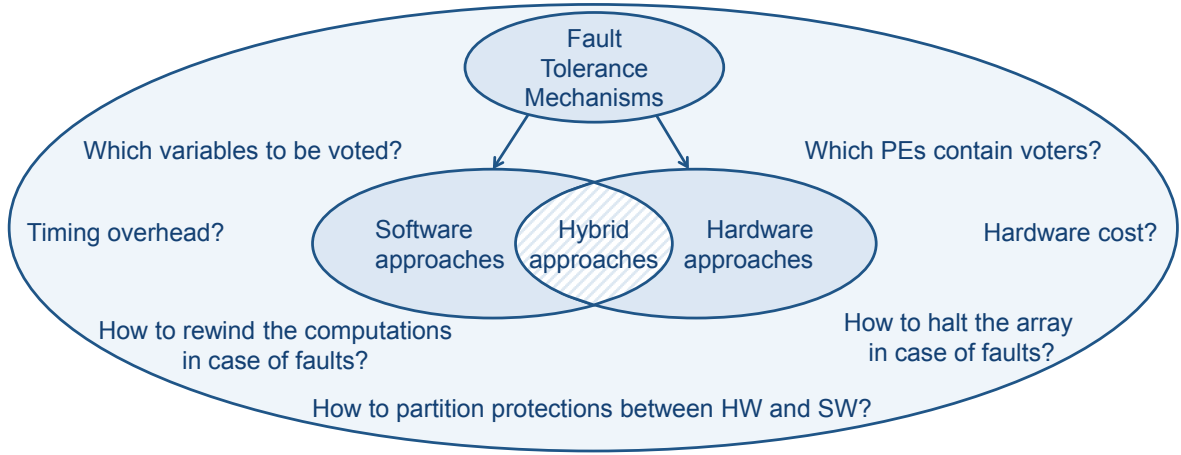

Fig. 3: Research perspectives for realizing fault-tolerant loop execution on TCPAs.

Another question raises up when detecting an error for example in case of DMR, what should be the reaction to the detected faults? Here, we need proper hardware facilities ensuring fail-safe halting the execution of the application affected by the error and suitable mechanisms to rewind back the execution to possible earlier points in time. We do not plan to implement any expensive check-pointing mechanisms inside PEs due to its high hardware cost, therefore, existing candidates for execution rewind could be either returning back to previous iterations (if all data dependencies and register life-times allow this), or to start of current input buffer or start of the input data volume (such as the start of a frame in case of the video processing applications). If a repetitive misbehavior is detected, then application might be migrated to another region on the processor array through a new invade request.

## VI. Concluding Remarks

Invasive computing is a resource-aware parallel computing paradigm giving applications the ability to expand their computations on available resources under functional and non-functional guarantees requested by application programers and based on the system status. This paper summarizes our hardware/software concepts for tightly coupled processor arrays to implement the concepts of invasive computing, i.e., enabling the reservation and release of computational resources during different phases of application executions. Here, each PE in the array is augmented with a dedicated hardware component called invasion controller, capable of acquiring PEs in either linear or rectangular regions in a distributed manner. The PE utilization information from invasion controllers are also directly used in order to have a adaptive power management control on individual PEs. Furthermore, we introduced our future research ideas toward guaranteeing fault-tolerant loop execution on TCPAs. We will investigate the possibility of on-demand usage of different levels of redundancies, e.g. dual/triple modular redundancy, based on application requests and the state of the system. In this regard, we will study of effects of voting/comparison operations on different loop intermediate variables and outputs. In order to make trade off between the flexibility of software-based voter and the low timing overhead of hardware-based ones, PEs will be equipped with specialized voter functional units that are capable of operating on any member of PE register files. Therefore, through a compiler characterization, we will be able to decide which parts of the code should be protected through software approaches and which part through hardware ones.

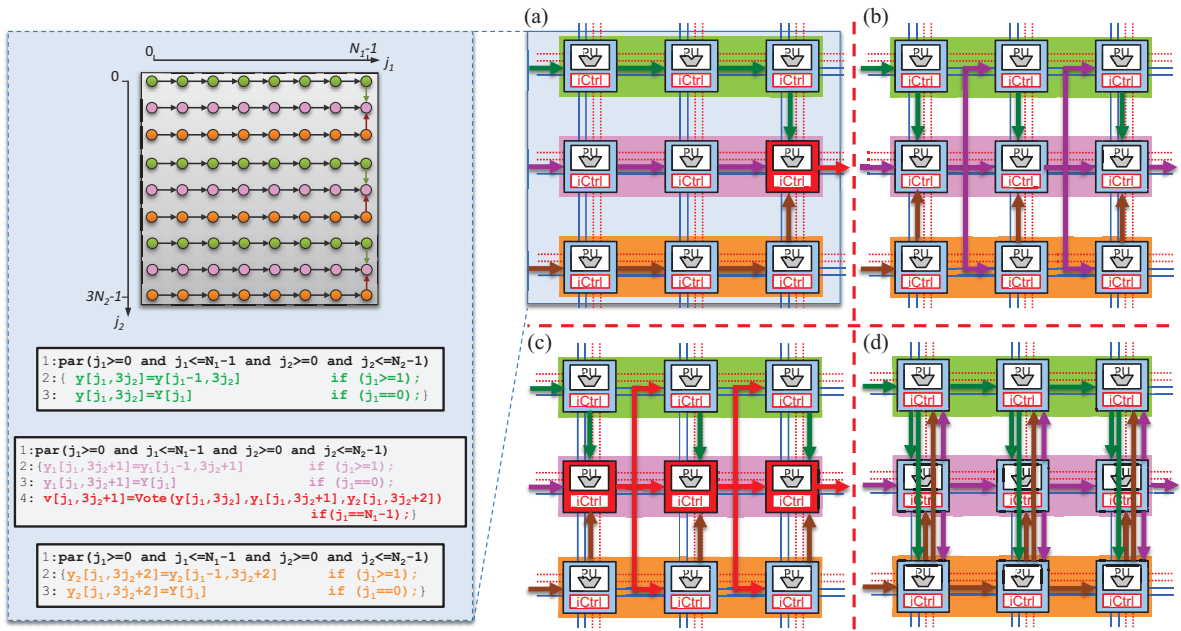

Fig. 4: Different voting mechanisms proposed for triple modular-redundant loop computations on TCPAs. The PEs colored in red contain *voter FUs*, voting on register file content at hardware level. (a) Hardware-based voting only on loop output variables that are mapped to border PEs (the red PE in the middle), for each processor array, the corresponding compiler generated codes (for TMR) and there iteration spaces are also shown. Disadvantage: Bad fault isolation/coverage. (b) Software-based voting on intermediate results, but performed by one of the replicas (middle row PEs) only. The voted output is then propagated to the other replicas for continued execution. (c) Hardware-based voting on intermediate results. (d) Software-based voting on all replicas.

ACKNOWLEDGMENTS

This work was supported by the German Research Foundation (DFG) as part of the Transregional Collaborative Research Centre "Invasive Computing" (SFB/TR 89).